\documentclass{pasj00}
\onecolumn
\begin{document}
\SetRunningHead{N.\ Sumitomo}
{Observational Appearance of Relativistic, Spherically Symmetric Massive Winds}
\Received{2006/00/00}
\Accepted{2006/00/00}

\title{Observational Appearance of Relativistic, Spherically Symmetric Massive Winds}

\author{Naoko \textsc{Sumitomo}, 
Shinji \textsc{Nishiyama}, 
Chizuru \textsc{Akizuki}
 \thanks{Present address: Center for Computational Physics, 
University of Tsukuba, Tennoudai 1-1-1, Tsukuba, Ibaraki, 305-8577}, 
Ken-ya \textsc{Watarai}
 \thanks{Research Fellow of the Japan Society for the Promotion of Science}, 
Jun \textsc{Fukue}} 
\affil{Astronomical Institute, Osaka Kyoiku University, 
Asahigaoka, Kashiwara, Osaka 582-8582}
\email{j069338@ex.osaka-kyoiku.ac.jp, fukue@cc.osaka-kyoiku.ac.jp}



\maketitle


\begin{abstract} 
The photon mean free path in a relativistically moving medium 
becomes long in the down-stream direction while short in the up-stream direction. 
As a result, the observed optical depth $\tau$ becomes small
in the downstream direction while large in the upstream direction.
Hence, if a relativistic spherical wind blows off,
the optical depth depends strongly on its speed and the angle
between the velocity and the line-of-sight. 
Abramowicz et al. (1991) examined such a relativistic wind, and found that
the shape of the photosphere at $\tau=1$ appears convex in the non-relativistic case, but concave for relativistic velocities. 
We further calculated the temperature distribution and luminosity of the photosphere both in the comoving and inertial frames. 
We found that the limb-darkening effect would strongly modified
in the relativistic regime.
We also found that luminosities of the photosphere becomes large as the wind speed increases due to the relativistic effects.
In addition, the luminosity in the inertial frame is higher than that in the comoving frame. 
These results suggest that the observed temperature and brightness in luminous objects may be overestimated when there are strong relativistic winds. 
\end{abstract}

\section{Introduction}

Recent observations have revealed the existence of highly luminous objects,
 such as ultraluminous X-ray sources (ULXs) in nearby galaxies,
 narrow line Seyfert 1 galaxies (NLS1s), and bright PG quasars. 
Bright objects are paid to attention because the outflow is observed in those objects. 
Quasar PG~1211+143 seems to have relatively high velocity outflow ($v \sim 0.1 c$)
 from their broad absorption line features observed by {\it XMM-Newton} (Pounds et al. 2003a). 
Quasar PG~0844+349 also shows several absorption features in its X-ray spectrum, and the outflow velocity is on the order of $\sim 0.2 c$ (Pounds et al. 2003b).   
Moreover, PKS~1549-79 is a luminous quasar-like active galactic nucleus,
 and contains relatively narrow permitted lines, highly blue-shifted [O\emissiontype{III}] lines,
 which are well-observed in NLS1s. 
The Very Large Telescope (VLT) observation of PKS~1549-79 shows outflow-like images,
 thus there is evidence that high mass accretion and warm outflow coexist in this object (Holt et al. 2006). 

Optical observations of broad absorption line (BAL) quasars also suggested the existence of not-collimated, relativistic outflows. 
Typical wind velocity estimated by their line width
 is about 10000--30000 km~s$^{-1}$.
The high column density of $N_{\rm H} \sim 10^{23-24} {\rm ~cm^{-1}}$
 suggests that large amounts of gas are moving around the central region of BAL quasars. 

King and Pounds (2003) recently suggested that highly optically thick, relativistic winds --- ``black hole winds'' --- blow off from the central part of accretion disks, and produce a large luminosity, which may be observed in quasar PG~1211+143
 and ULXs. 
If the mass-outflow rate of the wind increases,
 the optical depth of the wind exceeds unity,
 and eventually such a massive wind may form a ``photosphere'' like the sun. 
The location of the photosphere depends on the wind velocity and mass-outflow rate,
 then we may observe the light from an outflow rather than that from an accretion disk. 
Therefore, the determination of the location of the photosphere is quite important  from the observational viewpoint. 

Spherically symmetric, relativisitic winds have been investigated
by several researchers
(e.g., Castor 1972; Ruggles, Bath 1979; Mihalas 1980; Quinn, Paczy\'nski 1985; Turolla et al. 1986; Paczy\'nski 1990; Akizuki, Fukue 2007),
and their models have applied to neutron star winds and gamma-ray burst (GRB).
However, in these studies they focused on the dynamics of the relativistic outflow, and not concentrated on the observational implications. 
Abramowicz et al. (1991) first examined the observational appearance
of the relativistic, spherical winds,
using a simple model at a constant speed.
They found that
the shape of the photosphere appears convex in the non-relativistic case, but concave in a relativistic regime. 

%

Abramowicz et al. (1991) only considered the apparent shape
of the photosphere of the relativistic spherical winds.
Thus, in this paper, we further examine the relativistic spherical winds,
focusing our attention to the observational appearance,
such as an observed temperature distribution and an emergent luminosity,
 and give observational implications to several PG quasars and ULXs. 

In section 2, we briefly introduce the present wind model, and
describe the calculation method. 
Section 3 demonstrats the results, and
 the discussions and applications are presented in section 4. 
Final section is devoted to concluding remarks.

\section{Model and Calculation Method}

In this section
we describe the present wind model and calculation method.

\subsection{Wind Model}

Our present model is based on the model by Abramowicz et al. (1991). 
We briefly summarize the simple model at a constant velocity.

We assume that a spherically symmetric, relativistic wind blows off from a central object. 
As a central object,
we assume a non-rotating black hole (Schwarzschild black hole),
 and the Schwarzschild radius is defined by $r_{\rm g} = 2GM/c^2$,
 where $G$, $M$, and $c$ represent the gravitational constant, black hole mass,
 and the speed of light, respectively.
We use the spherical coordinates $(R, \theta, \varphi)$ and
the cylindrical coordinate $(r, \varphi, z)$,
whose $z$-axis is along the line-of-sight  (see figure 1). 

From continuity equation, 
the rest mass density $\rho_{\rm 0}$ measured in the comoving frame
varies as  
\begin{equation}
  \rho_{\rm 0} = \left ( \frac{\dot{M}}{4 \pi v \gamma} \right) R^{-2},
\end{equation}
 where 
$\dot{M}$ is the mass-loss rate, 
$R=\sqrt{r^2 +z^2}$ is a distance from the central object, 
and $\gamma$ is the Lorentz factor of the wind expressed as 
\begin{equation}
   \gamma \equiv (1-\beta^2)^{-1/2},~~ \beta \equiv \frac{v}{c}.  
\end{equation} 
Quantities with subscript ``0'' refer to physical quantities
 measured in the comoving frame. 

In the present simple model,
the mass-loss rate $\dot{M}$ and wind velocity $v$ are assumed to be constant. 

\begin{figure}
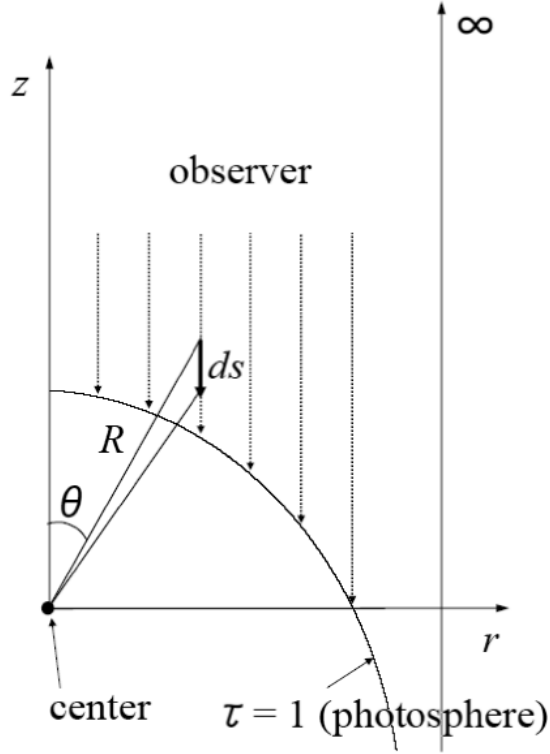

  \begin{center}
  \FigureFile(80mm,80mm){figure1.eps}
@\end{center}
\caption{Schematic picture of the present calculation.
The spherical wind is assumed to blow off from the origin
at a relativistic speed.
The observer is located at infinity in the $z$-direction.}
\end{figure}

\subsection{Photosphere of the Wind}

We define an ``apparent photosphere'' of the wind as the surface, where
 the optical depth $\tau$ measured from an observer becomes unity.  
Schematic picture  of our present calculation is presented in figure 1.

Lets us consider a small distance $ds$ along the light path. 
Due to the relativistic effect,
the mean free path of photons in the fixed frame, $\ell$,
is related to that in the comoving frame, $\ell_0$, by
\begin{equation}
    \ell = \frac{1}{\gamma (1- \beta \cos \theta)} \ell_0,
\end{equation}
where $\theta$ is the viewing angle measured from the $z$-axis.
Then,  the optical depth in the fixed frame is given by 
\begin{equation}
  d\tau = \gamma (1- \beta \cos \theta) \kappa \rho _{\rm 0} ds,
  \label{eq:dtau}
\end{equation}
where the opacity $\kappa$ is assumed to be electron scattering
(Abramowicz et al. 1991). 
Thus, the optical depth strongly depends on the viewing angle $\theta$ 
as well as the flow speed $v$.
It is obvious that the optical depth is smallest 
in the downstream direction at $\theta=0$, 
where the photons move in the same direction with the fluid,
while it becomes largest in the upstream direction at $\theta=\pi$,
where the photons move in the opposite direction to the fluid. 

From equation (\ref{eq:dtau}), the integrated optical depth $\tau_{\rm ph}$ from an observer at infinity is calculated as 
\begin{equation}
\tau_{\rm ph} = \int_{z_{\rm ph}}^\infty \gamma (1- \beta \cos \theta) \kappa \rho _{\rm 0} ds=1,
\label{eq:tauph}
\end{equation}
where ${z_{\rm ph}}$ is location of the apparent photosphere from the equatorial plane.
Abramowicz et al. (1991) showed that the photosphere of a highly relativistic wind 
is much closer to the source, roughly by a factor $\gamma^2$. 
Although the non-relativistic and moderately relativistic winds have convex photospheres, the photospheres of relativistic wind becomes concave for $\beta > 2/3$ (see figure 2).

\subsection{Temperature Distribution and Luminosity}

In the present model,
we assume that the spherical wind expands adiabatically.
Then, the temperature $T_0$ of the wind gas in the comoving frame
varies as
$
   T_0 \propto \rho_0^{\Gamma - 1} \propto \rho_0^{1/3} \propto R^{-2/3},
$
where $\Gamma$ is the ratio of specific heats,
and set to be $4/3$ for a radiation pressure dominant regime.
Hence, the temperature distribution in the comoving frame is
\begin{equation}
   \frac{T_0}{T_{\rm c}} = \left( \frac{R}{R_{\rm c}} \right)^{-2/3},
\end{equation}
where
$T_{\rm c}$ is the central temperature at $R=R_{\rm c}$.

Furthermore, the observed temperature $T$ 
in the observer's frame is expressed by
\begin{equation}
T = \frac{1}{1+z} T_0 = \frac{1}{\gamma (1- \beta \cos \theta)}T_{\rm 0},
\end{equation} 
where $z$ is the redshift via longitudinal and transverse Doppler effects.
Using this observed temperature,
we can obtain the observed luminosity by 
\begin{equation}
L = \int_{r_{\rm in}} ^{r_{\rm out}} 2\pi rdr\times F dr,
\end{equation} 
where $F$ is the observed flux,  $F = \sigma T^4$ when we assume the blackbody radiation, $\sigma$ being the Stefan-Boltzmann constant.
This integration has been performed at $z \to \infty$ in the observer's frame,
 thus the effect of curvature of the photosphere is automatically included. 

\section{Results}

We determined the photosphere for various $\beta$ via equation (\ref{eq:tauph}), 
 and obtained the temperature and luminosity on the surface of the photosphere in the comoving and inertial frames. 
In the present calculation,
we bear in mind the case for active galactic nuclei,
although the present results are also important for black-hole binaries.
Hence,
the black hole mass and temperature at the central region are fixed as
 $M=10^7 M_{\odot}$ and $T_{\rm c} =10^7 K$, respectively. 
The input parameters are then the velocity $\beta$ and 
the normalized mass-loss rate $\dot{m}$ of wind, where
the mass-loss rate is normalized by the critical rate,
 $\dot{m}=\dot{M}/\dot{M}_{\rm crit}=\dot{M}/(L_{\rm E}/c^2)$, 
$L_{\rm E}$ being the Eddington luminosity.

\subsection{Location of the Photosphere}

Figure 2 shows the location of the apparent photosphere seen by the observer at infinity in the $z$-direction for various wind velocities.  
In the low speed regime,
the photosphere near the $z$ axis is close to the center,
 while the photosphere far away from the  $z$ axis is far from the center. 
This is the usual limb-darkening effect of the spherically expanding wind.

In the high speed regime, on the other hand,
the shape of the apparent photophere changes,
because the optical depth depends on the angle $\theta$ and the wind velocity $v$.
In particular, when the wind blows off at highly relativistic speed ($\beta \geq 0.8$),
 the photosphere looks like a concave shape.

Our results consistent with the analytical results by Abramowicz et al. (1991), 
 but we note that the units of our coordinate is the Schwarzschild radius $r_{\rm g}$. 

\begin{figure}
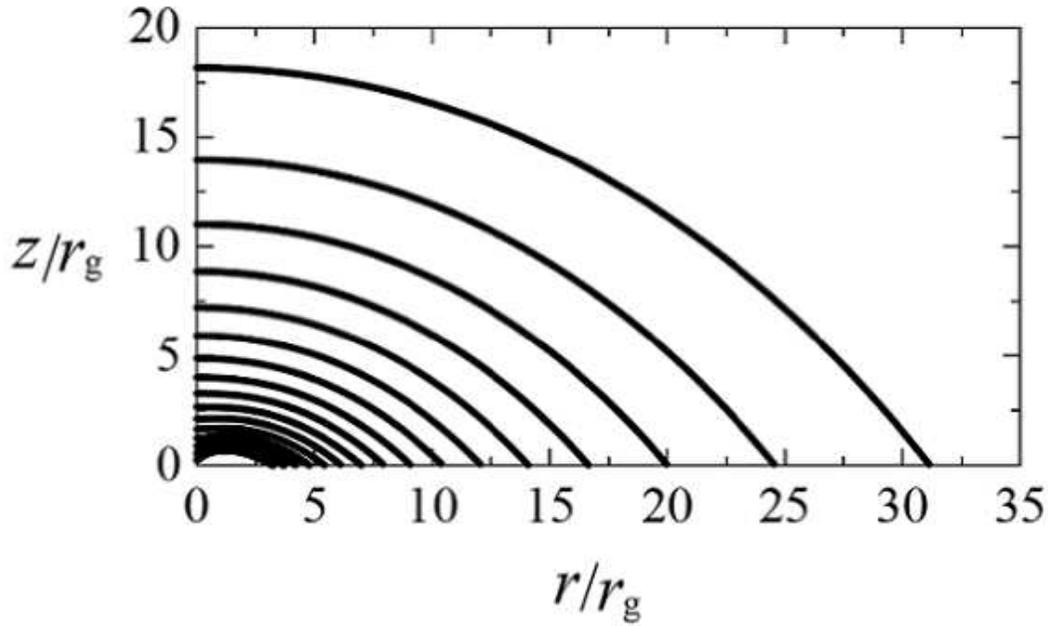

  \begin{center}
  \FigureFile(140mm,140mm){figure2.eps}
  \end{center}
\caption
{
Location of the photosphere for various wind velocity $\beta$.
The wind velocity is varied from 0.95 (central region) to 0.20 (outer region) in steps of 0.05. 
The units of the $r$- and $z$-axis is the Schwarzschild radius $r_{\rm g}$.
}
\end{figure}

\subsection{Temperature Distribution}

In figure 3 we show the temperature distributions in the comoving and inertial frames, respectively, viewed by an observer at infinity
in the direction of the $z$-axis. 
In general,
the wind photosphere looks brightest at the central part, and the surroundings are gradually dim as increasing radius.
This is due to the limb darkening effect seen in the usual spherical wind.

In the relativistic wind considered here, however,
this limb-darkening effect is remarkably enhanced.
This is
due in part to the relativistic Doppler and aberration effects,
and due in part to the fact that
the observed photosphere shrinks as the velocity increases and
we can see deep inside the wind.

Comparing the temperature distributions in the comoving and inertial frames,
we see that
the central temperature in the inertial frame is higher than that in the comoving frame. 
This is just the relativistic Doppler and aberration effects.
That is, the observed temperature increases as $\theta$ approaches zero
 because of the longitudal Doppler effect (highly beamed emission).
This effect also becomes remarkable as the velocity increases. 

\begin{figure}
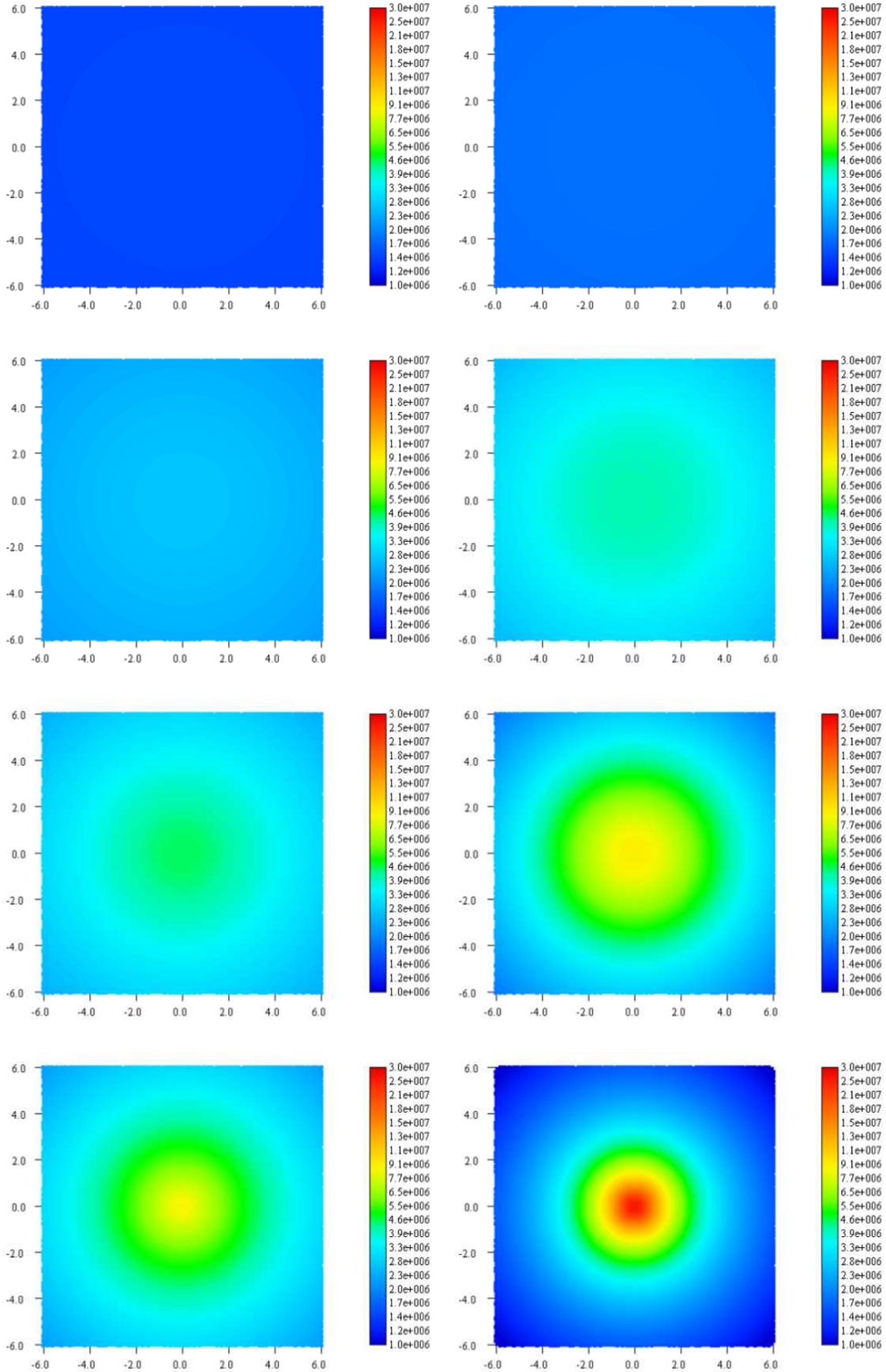

  \begin{center}
  \FigureFile(75mm,75mm){figure3a.eps}
  \FigureFile(75mm,75mm){figure3aa.eps}
  \end{center} 
\begin{center}
  \FigureFile(75mm,75mm){figure3b.eps}
  \FigureFile(75mm,75mm){figure3bb.eps}
\end{center} 
\begin{center}
  \FigureFile(75mm,75mm){figure3c.eps}
  \FigureFile(75mm,75mm){figure3cc.eps}
\end{center} 
\begin{center}
  \FigureFile(75mm,75mm){figure3d.eps}
  \FigureFile(75mm,75mm){figure3dd.eps}
\end{center} 
\caption{
Temperature distribution at the apparent photosphere
viewed by an observer at infinity in the $z$-direction.
The left and right panels show the temperatures in the comoving and fixed (observer's) frames, respectively.
The wind velocity $\beta$ is varied as 0.2, 0.4, 0.6, and 0.8 from top to bottom in both panels. 
}
\end{figure}

\subsection{Apparent Luminosity}

In figure 4
we show the luminosities of the relativistic winds
as a function of the wind velocity $\beta$
for several mass-loss rates $\dot{m}$.
Solid curves represent the observed luminosity in the inertial frame,
 and the dashed curves show the comoving luminosity.

As is seen in figure 4,
the wind luminosity increases as the velocity increases,
while it decreases as the mass-loss rate increases.
In addition,
the luminosity observed in the inertial frame is higher than
that in the comoving frame.

In the present calculation, the location of the photosphere strongly depends on the density of the wind. 
Namely, as the mass-loss rate increases, the wind density increases, and the location of the photosphere expands to the outer region. 
Since the temperature of the wind in the outer region is lower than that in the inner region, 
the luminosity decreases as the mass-loss rate increases. 

%
Moreover, by the effect of Doppler beaming, the observed luminosity increases
as the velocity increases.
It is emphasized that the comoving luminosity is enhanced about 130 percent  for $\beta=0.2$, but the amplification is about one order for $\beta=0.9$. 
These facts suggest that there is a possibility of overestimation of the observed luminosities for relativistic outflow objects.

\begin{figure}
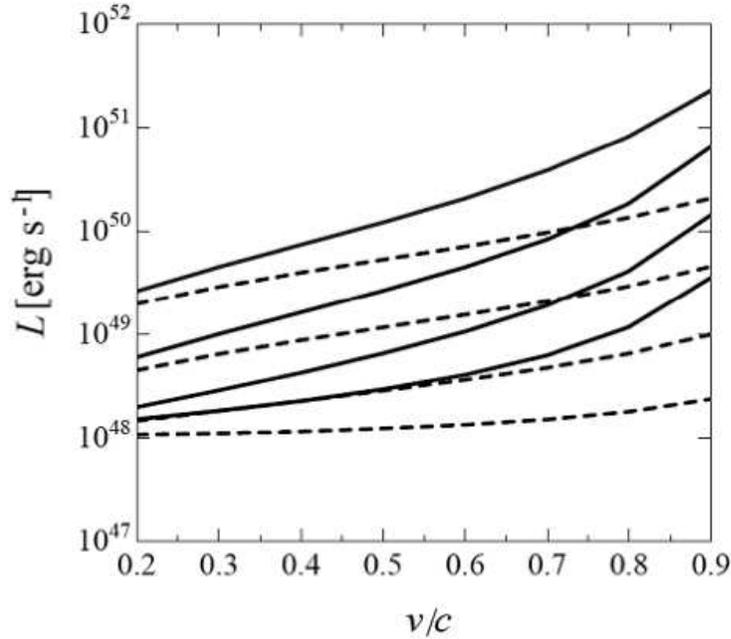

  \begin{center}
  \FigureFile(100mm,100mm){figure4.eps}
  \end{center}
\caption{
Luminosity of the relativistic wind as a function of the wind velocity
for several mass-loss rates.
Solid curves represent the observed luminosity in the inertial frame,
 and the dashed curves show the comoving luminosity.
The mass-loss rate ${\dot{m}}$ are 10, 100, 1000, and 10000 from top to bottom. 
}
\end{figure}


\section{Discussions}

\subsection{Observational Importance of the Apparent Photosphere}

The observed shape of the photosphere is asymmetric
 in spite of  spherically symmetric winds.
This is due to the limb-darkening effect. 
This nature does not depend on the observer's direction.
Due to the optical depth effect,
we could see deeper inside the wind, as the velocity increases.
In addition, the luminosity in the observer's frame is remarkably enhanced by relativistic beaming effects along the observer's direction. 
These two effects mainly work as the luminosity enhancement of the relativistic outflow. 

This fact shows the possibility of overestimation of the temperature and the luminosity 
of the object that is supposed for wind to blow off at a relativistic speed.
When the wind velocity is fast or the mass-loss rate is small, 
the radius on a bright photosphere becomes small,
 and we observe higher temperature and higher luminosity.
%
The radius on a bright photosphere increases, when the speed is small or the mass-loss rate increase. 

If the wind blows off from the luminous accretion disk surrounding the central object, the disk is concealed by the massive wind, there is a possibility of observing 
not disk but wind (cf. Nishiyama et al. 2007). 
There is possibility that the estimate of luminosity and black hole mass different from those actual values.

\subsection{Relation to BAL, PG Quasars} 

The quasar PG 1211+143 is a strong candidate having an optically thick wind driven
 by radiation pressure, and its wind velocity is roughly estimated to be $0.1 c$, using X-ray observation by {\it XMM-Newton} (Pounds et al. 2003a). 
Typical velocity of the outflow in BAL quasars also shows sub-relativistic velocities
 ($\sim 0.1-0.3 c$). 
These luminous quasars are likely to have sub-relativistic outflow,
 thus acceralation mechanism of the wind may be same. 
It is known that the radiation-pressure driven wind is difficult to be acceralated to highly relativistic speed ($\beta \sim 0.9$), but possible to be acceralated up to mildly relativistic speed ($\beta \sim 0.1-0.3$)  (e.g., Icke 1980; Tajima, Fukue 1997; Watarai, Fukue 1999). 
Recent 2D radiation hydrodynamical simulations also showed massive, optically thick winds from a luminous accretion flow, but their wind velocities are still  subrelativistic of $0.1-0.3 c$ (e.g., Proga 2003; Ohsuga et al. 2005). 
Unfortunately, a sample number of PG quasars is insufficient for a statistical argument. 
Moreover, absorption lines observed in PG~0844+349 are highly suspectable because 
 re-analysis of the same data by other group did not confirm the earlier results  (Brinkman et al. 2006). 
Thus we hope further reliable detections of the broad absorption lines in BAL quasars to confirm the existence of mildly relativistic outflows.

\section{Concluding Remarks} 

In this paper, we examined the appearance of relativistic, spherically symmetric wind from the observational point of view.
As for the shape of the apparent photosphere of massive winds,
we confirmed the results of Abramowicz et al. (1991).
We further calculated the temperature distribution and luminosity of the photosphere both in the comoving and inertial frames. 
We found that the limb-darkening effect would strongly modified
in the relativistic regime.
We also found that luminosities of the photosphere becomes large as the wind speed increases due to the relativistic effects.
In addition, the luminosity in the inertial frame is higher than that in the comoving frame. 
In particular, the luminosity in the observer's frame
 is one order of magnitude higher than that in the comoving frame for highly relativistic regimes. 
We suggest that
 if the observed luminosity is used for the evaluation of the black hole mass, 
 then the derived black hole mass will be overestimated. 

In order to compare with observational data,
 we need more strict treatments of a wind model,
 e.g., the effect of general relativity, radiative energy loss in the wind,
 compton processes, acceralation by radiation pressure, etc. 
However, the aim of this paper is to show the possibility of the formation of the photosphere. 
Here, we explicitly show the formation of the photosphere
 in an optically thick wind using a simple spherical wind model.

Strictly speaking, the observed temperature should be evaluated from the temperature on the surface, where the effective optical depth equals to unity,
$\tau_{\rm eff}=\sqrt{\tau_{\rm ff} \tau_{\rm tot}}=\sqrt{\tau_{\rm ff} (\tau_{\rm ff}+\tau_{\rm es})}=1$,
$\tau_{\rm ff}$ and $\tau_{\rm es}$ being the free-free and electron scattering opacities, respectively. 
In a high temperature plasma,
the effective optical depth is often smaller
that the total optical depth,
and the gas becomes scattering dominant.
In such a scattering dominated plasma,
the emergent spectrum is not a simple blackbody
but a modified blackbody
(e.g., Rybicki and Lightmann 1979).
In addition,
we have used the Thomson cross section for electron scattering.
When the center-of-mass energy of scattering becomes relativistic
($h\nu \sim 100$~keV),
we should use the Klein-Nishina cross section,
which reduces the effective cross section.
In such a highly relativistic regime,
a wind will be much more transparent.
These effects are also left as future problems.


\vspace{10mm}

This work was supported in part by the Grant-in-Aid 
for JSPS fellows (16004706 KW) and
for Scientific Research of the
Ministry of Education, Culture, Sports, Science, and Technology
(18540240 JF).


\end {document}